\documentclass[twocolumn,pre,aps]{revtex4}
\usepackage{graphicx}
\usepackage{dcolumn}
\usepackage{bm}
\usepackage{amsmath}
\begin{document}
\newcommand{\be}{\begin{equation}}
\newcommand{\ee}{\end{equation}}
\newcommand{\ba}{\begin{eqnarray}}
\newcommand{\ea}{\end{eqnarray}}

\title{Properties of Foreshocks and Aftershocks of the Non-Conservative SOC
Olami-Feder-Christensen Model: Triggered or Critical Earthquakes?}
\author{Agn\`es Helmstetter}
\affiliation{ Institute of Geophysics and Planetary Physics, 
University of California, Los Angeles, California 90095-1567.}
\author{Stefan Hergarten}
\affiliation{Geodynamics-Physics of the Lithosphere, University of Bonn, Germany}
\author{Didier Sornette}
\affiliation{Department of Earth and Space Sciences and Institute of
Geophysics and Planetary Physics, University of California, Los Angeles,
       California 90095-1567 and
     Laboratoire de Physique de la Mati\`{e}re Condens\'{e}e, CNRS UMR 6622
Universit\'{e} de Nice-Sophia Antipolis, Parc Valrose, 06108 Nice, France}

\begin{abstract}
Following Hergarten and Neugebauer \cite{StephanPRL} who discovered 
aftershock and foreshock sequences in the Olami-Feder-Christensen (OFC) 
discrete block-spring earthquake model, we investigate to what degree 
the simple toppling mechanism of this model
is sufficient to account for 
the properties of earthquake clustering in time and space.
Our main finding is that synthetic catalogs generated by the OFC model
share practically all properties of real seismicity at a qualitative
level, with however significant quantitative differences. We find that
OFC catalogs can be in large part described by the concept of triggered
seismicity but 
the properties of foreshocks depend on the mainshock magnitude,
in qualitative agreement with the critical earthquake model and in
disagreement with simple models of triggered seismicity such as the 
Epidemic Type Aftershock Sequence (ETAS) model \cite{Ogata88}.
Many other features of OFC catalogs can be reproduced with the
ETAS model with a weaker clustering than real seismicity, i.e. for
a very small average number of triggered earthquakes of first generation per
mother-earthquake. Our study also evidences the large biases
stemming for the constraints used for defining foreshocks and aftershocks.
\end{abstract} 

\maketitle
\vskip 0.5cm

\section{Introduction}
Describing and modeling the space-time organization of seismicity and
understanding the underlying physical mechanisms to quantify
the limits of predictability remain important open challenges.
Inspired by statistical regularities such as the Gutenberg-Richter
\cite{GR44} and the Omori \cite{Omori}
laws, a wealth of mechanisms and models have been proposed.
New classes of models inspired or derived from statistical physics
accompanied and followed the proposition, repeated several times under
various forms in the last 25 years, that
the space-time organization of seismicity is similar to the behavior
of systems made of elements interacting at many scales
that exhibit collective behavior such as in critical phase transitions.
This led to the concepts of the critical earthquake, of self-organized
criticality, and more generally of the seismogenic crust
as a self-organized complex system requiring a so-called system approach.

Our purpose here is to study in depth maybe the simplest model
of the class of self-organized critical models that exhibit
a phenomenology resembling real seismicity, the so-called OFC sandpile model.
Real seismicity is usually divided into three major classes
of earthquakes, foreshocks, mainshocks, and
aftershocks. Many different mechanisms have been discussed to
explain these three classes. The OFC model
uses only one simple local interaction between discrete fault
elements but nevertheless exhibits a sufficiently rich behavior
that these three classes of events are observed \cite{StephanPRL}.
The motivation of our present work is thus to study
the main characteristics of foreshocks and aftershocks in the OFC
model, and to understand the mechanisms responsible for earthquake
clustering in the OFC model.
For this, we will
interpret our analysis of the OFC catalogs in the light of two
end-member models representing two opposite views of seismicity.
The first one, mentioned above,
is the critical earthquake model, which views a mainshock as the special
outcome of a global self-organized build-up occurring at smaller scales.
The second end-member is called the ETAS (Epidemic-Type Aftershock Sequence) 
model and is nothing but a phenomenological construction based on
the well-documented empirical Gutenberg-Richter law, the Omori law and
the aftershock productivity law. In the ETAS model, all earthquakes are put
on the same footing, any earthquake can trigger other
earthquakes and be triggered by
other earthquakes and the distinction between foreshocks, mainshocks
and aftershocks is only for convenience and does not reflect
any genuine difference. The ETAS model can thus be considered
as a realistic statistical null-hypothesis of seismicity.

The plan of our paper is the following. Section \ref{OFCmodel} presents 
the OFC model, section \ref{nghjej} summarizes the phenomenology of real 
seismicity, and section \ref{models} presents the critical earthquake model 
and the ETAS model. Section \ref{result} presents the results obtained for 
the OFC model, that are compared with real seismicity and with the two
reference models (ETAS and  critical earthquake models).
We discuss in section \ref{discu} possible mechanisms for 
foreshocks and aftershocks in the OFC model.
Section \ref{conclu} concludes.

\section{The Olami-Feder-Christensen (OFC) model}
\label{OFCmodel}

The Olami-Feder-Christensen (OFC) model
\cite{Olami} is defined on a discrete system of blocks or of
fault elements on a square lattice, each carrying a force.
The force $F_i$ of a given element $i$ that exceeds a
fixed threshold $F_{\rm c}$ (taken equal to $1$
without loss of generality) relaxes to zero. Such a toppling increments 
the forces on its nearest neighbors by a pulse which is
$\alpha\:(\alpha\le1/4)$ times the force $F_i$ of the unstable element:
\begin{equation}
F_i\ge F_{\rm c} \Rightarrow \left\{\begin{array}{lll} F_i &\rightarrow &0 \\
F_{nn} &\rightarrow &F_{nn}+\alpha F_i~.
\end{array}\right. \label{OFC}
\end{equation}
This loading to nearest-neighbors can in turn destabilize
these sites, creating an avalanche.
Between events, all $F_i$'s increase at the same constant
rate, mimicking a uniform tectonic loading.
The OFC model can be obtained as a sandpile analogue of block-spring
models \cite{BurridgeKnopoff}, with the interpretation
\be
\alpha = {1 \over n_i +k}
\ee
where $n_i$ is the actual number of neighbors of site $i$.
$n_i$ is always 4 in the case of a square lattice with rigid-frame boundaries.
For free boundary conditions used in this study
$n_i=4$ in the bulk and $n_i=3$ (resp. $2$) at the boundaries (resp. corners).
The symbol $k$ denotes the elastic constant of the upper leaf springs, measured
relatively to that of the other springs between blocks.
The OFC model is conservative for $k=0$ for which
$\alpha =0.25$ and is non-conservative for $k>0$ for which $\alpha < 0.25$.
In the following, we will compare results obtained for $k=0.5, 1, 2$ and $4$,
that is for $\alpha=0.222, 0.2, 0.167$ and $0.125$.

With open or rigid boundary conditions, this model seems to
show self-organized criticality (SOC)
\cite{Bak,jensenSOC,critbook} even in the non-conservative case $\alpha < 0.25$.
SOC is the spontaneous
convergence of the dynamics to a statistically stationary state characterized
by a time-independent power law distribution of avalanches. The size
of an avalanche is taken to be the spanned area $s$.
The underlying mechanism for SOC seems to
be the invasion of
the interior by a region spreading from the
boundaries, self-organized by the synchronization or
phase-locking forces between the individual elements \cite{Others5}.

Long-term correlation between large events have been documented but,
only very recently, Ref.~\cite{StephanPRL} found the occurrence of
genuine sequences of foreshocks and aftershocks that bear similarities
with real earthquake catalogs. This discovery is
interesting because it suggests that a unique mechanism is sufficient
to produce a Gutenberg-Richter-like distribution as well as
realistically looking foreshocks and
aftershocks, without the need for viscous crust relaxation or other
mechanisms. Similarly, Ref.~\cite{Huangetla98} found critical precursory
activity and aftershock sequences in a sandpile model. However, the 
precursors and aftershocks resulted from the interplay between the 
built-in hierarchy of domains and a conservative sandpile dynamics. 
The remarkable observation of Ref.~\cite{StephanPRL} is that such a 
hierarchy is not needed for foreshocks and aftershocks to occur, when 
the sandpile dynamics is dissipative. However, a hierarchy of faults may be
needed to obtain a larger number of triggered events than found 
for the OFC model which is more compatible with real seismicity.
Our goal here is to investigate 
in details the properties of the foreshocks and aftershocks in such a 
situation, that is, in the OFC model.

Our simulations presented below are performed in two-dimensional
square lattices $L \times L$ with $L=512, 1024$ and $2048$.
Let us give a correspondence between time and space units in the OFC
model and in the real seismicity.
If we consider that the lattice of $L=1024$ blocks
represents a fault of $100 \times 100$ km (we neglect
the asymmetrical aspect ratio of real faults), which is able to produce 
an earthquake with magnitude $M=8$,
the minimum earthquake of size $s=1$ generated by the OFC model 
has a length of $\approx 0.1$ km, corresponding to a magnitude 2
earthquake. Ref.~\cite{StephanPRL} documents that the typical waiting time between
two events involving at least $1000$ blocks is $10^{-2}$ in the time unit of the model. 
This event of size $s=1000$ corresponds to a rupture length of
$\approx 3$ km or a magnitude $M \approx 5$ earthquake. 
There are about three events per year with magnitude $M \geq 5$ in Southern California.
This gives the correspondence between time units $\rm time_{\rm OFC}$ in the
OFC model and in real seismicity:
\be
{\rm time}_{\rm OFC} = 10^{-2}    ~~~~\Longleftrightarrow
~~~~{\rm time}_{\rm real} = 100 ~{\rm days}~.
\label{jfkkfw}
\ee
Ref.~\cite{StephanPRL} describe a particularly active aftershock sequence following
an event moving $256^2$ blocks (magnitude $\approx 7$) lasting
${\rm time}_{\rm OFC} = 4 \cdot 10^{-6}$ containing about 2500 events.
According to (\ref{jfkkfw}), this corresponds to an aftershock
sequence lasting only one hour.
The OFC model clearly does not have the time and space scales right to describe the
seismicity in California or in other regions of the world. 
The number of aftershocks is much smaller in
the OFC model than in California, and aftershocks occur at very short
times after the mainshock. 
In the OFC model, aftershocks start at  $t_{OFC}=10^{-14}$
after the mainshock (resulting from the numerical precision of the simulation).
In real seismicity, 
aftershocks are only observed after a few minutes following large $M=7$
mainshocks, due to the duration of the mainshock rupture and to the saturation 
of the seismic network. In real seismicity, aftershocks are also observed
over much longer times, from months to years. The aftershock sequences in the
OFC model resemble perhaps more those of deep earthquakes, which have 
few aftershocks \cite{Tonga,omorideep}.

\section{Phenomenology of real seismicity}
\label{nghjej}

The empirical properties of real seismicity discussed in this paper are
the following.

\begin{enumerate}
\item The Gutenberg-Richter (GR) law \cite{GR44}
states that the density distribution function $P(m)$
of earthquake magnitudes $m$ is
\be
P_0(m) dm = b~ \ln(10) ~ 10^{-b (m-m_d)}~dm~, \label{gojfwo}
\ee
with a $b$-value usually close to $1$.
$m_d$ is usually a lower bound magnitude of detection, such that
$P_0(m)$ is normalized to $1$ by summing over all magnitudes
above $m_d$.

The qualifying property of SOC in the OFC model is the existence of a 
GR-like distribution
of avalanches sizes. There are several measures of sizes. If we take the size
defined as the area $s$ spanned by an avalanche, the distribution of 
event sizes
is exactly given by (\ref{gojfwo}), where the magnitude $m$ is defined by
\be
m = \log_{10}(s)~.  \label{mgmlee}
\ee
The number of topplings is not exactly equal to $s$ since a site
can topple more than once in a given avalanche, possibly being reloaded
during its development. However, the difference is negligible for our purpose.
No multiple relaxations were observed for $k\geq2$.  
For $k = 1$, less than 1 multiple relaxation per 100,000 earthquakes was found. 

\item The (modified) Omori law \cite{Omori,KK78,Utsu} describes the
decay of the seismicity rate triggered
by a mainshock with the time $t$ since the time $t_c$ of the mainshock
\be
N_a(t) = {K_a  \over (t-t_c+c)^{p_a}} ~,
\label{first}
\ee
with an Omori exponent $p_a$ close to $1$. This decay law can be
detected over time scales spanning from weeks up to decades.
The time shift constant $c$ ensures a finite seismicity rate just
after the mainshock and is often of the order of minutes.

\item The inverse Omori law \cite{Papazachos,KK78,JM79}
describes the average increase of seismicity observed before a
mainshock and is like (\ref{first}) with $t-t_c$ replaced by $t_c-t$:
\be
N_f(t) = {K_f  \over (t_c-t+c)^{p_f}} ~,
\label{firstsad}
\ee
with the inverse Omori exponent $p_f$ usually close to or slightly smaller
than $p_a$ \cite{Helmfore}.

In contrast with the direct Omori law
which can be clearly observed in a single sequence following a mainshock,
the inverse Omori law can only be found by stacking many foreshock sequences
because there are huge fluctuations of the rate of foreshocks for individual 
sequences preventing the detection of any acceleration of seismicity.
The well-defined acceleration (\ref{firstsad}) of the seismicity preceding 
mainshocks emerges only when averaging over many foreshock sequences 
\cite{HSforedata03}.
\item {\it The productivity law} documents the growth of the number of
triggered aftershocks as a function of the magnitude $m$ of the mainshock:
\be
N_a(m) \sim ~10^{\alpha_a m}~,  \label{formrho}
\ee
where the exponent  $\alpha_a$ is usually found in the range
$0.7-1$ (see \cite{HelmPRL} and references therein).
This value of the exponent $\alpha_a$ may reveal a fractal spatial 
distribution of aftershocks \cite{HelmPRL}.

\item {\it Aftershock diffusion}.
Several studies have reported ``aftershock diffusion,'' the phenomenon of
expansion or migration of aftershock zone with time \cite{Tajima1,Marsan2}.
However, the present state of knowledge on aftershock diffusion
is confusing because contradictory results have
been obtained, some showing almost systematic diffusion whatever the
tectonic setting and in many areas in the world,
while others do not find evidences for aftershock diffusion
\cite{Shaw,HOS03}.
The shift in time from the dominance of the aftershock activity
clustered around the mainshock
at short times after the mainshock to the delocalized
background activity at large times may give rise to
an apparent diffusion of the seismicity rate when using
standard quantifiers of diffusion processes \cite{HOS03}.

\item {\it Foreshock migration}.
Foreshock migration towards the mainshock as time increases up to
the time of the mainshock has also been documented
\cite{KK78,VS81,HSforedata03} but may be due to an artifact of the background
activity, which dominates the catalog at long times and distances from
the mainshock \cite{HSforedata03}. Indeed, by an argument
symmetrical to that for aftershocks,
the shift in time from the dominance of the
background activity at large times before the mainshock to that of
the foreshock activity clustered
around the mainshock at times just before it may be taken as
an apparent inverse diffusion of the seismicity rate.

\item {\it The average distance $R_a$ between aftershocks and the mainshock
rupture epicenter} has been found to be proportional to the
rupture size of the mainshock, leading to the following scaling law
\cite{K02}
\be
R_a \sim 10^{0.5 m} \sim s^{0.5}~,
\label{mjglsacl}
\ee
relating $R_a$ and the mainshock magnitude $m$ or the mainshock
rupture surface $s$.
A similar law is believed to hold for the average distance $R_f$
between foreshocks and the mainshock \cite{KBMalinovskaya,Bowman}

\item {\it Foreshock magnitude distribution.}  
Many studies have found that the apparent $b$-value of the
magnitude distribution of foreshocks is smaller than that of
the magnitude distribution of the background seismicity and of
aftershocks (see \cite{Helmfore} and references therein).

\item {\it Number of foreshocks and aftershocks per mainshock.}
Foreshocks are less frequent than aftershocks \cite{KK76,KK78,JM79}.
The ratio of foreshock to aftershock numbers is in the range 2-4 for
$M=5-7$ mainshocks, when selecting foreshocks and aftershocks at a distance 
in the range $R=50-500$ km from the mainshock and for a time 
in the range $T=10-100$ days before or 
after the mainshock \cite{KK76,KK78,JM79,VS81,Shaw}. 

\end{enumerate}

\section{End-member models of seismicity: ETAS and critical earthquake 
models \label{models}}

\subsection{The ETAS model \label{gjkdks}}

The ETAS (Epidemic-Type Aftershock Sequence) model was introduced
in \cite{Ogata88} and in \cite{KK817} (in a slightly different form).
Contrary to what its name may imply, it is not only a model of aftershocks
but a general model of seismicity.

This model of multiple cascades of earthquake triggering avoids the division 
between foreshocks, mainshocks and aftershocks because it uses the same laws 
to describe all earthquakes.
Because of its simplicity, it is natural to consider it as a null hypothesis 
to explain the OFC catalogs and real data. 
Its choice as a reference is
also natural because it is a nothing but a branching model of earthquake
interactions and can thus be considered as a mean field approximation of
more complex interaction processes. Branching processes
can also be considered as natural mean field approximations of SOC models
and in particular of the OFC model \cite{Alstrom1} (see also
chapter 15 in \cite{critbook}). The approximation consists
usually in the fact that branching models neglect the loops
occurring in the cascade characterizing a given avalanche.
Note that standard branching models study the development of a single
avalanche while ETAS describe a catalog of earthquakes.

The ETAS model uses three of the above empirical laws as direct inputs
(Gutenberg-Richter law (\ref{gojfwo}), Omori's law (\ref{first}) 
and aftershock productivity law (\ref{formrho})).
In the ETAS model, a main event of magnitude $m$ triggers its own
primary aftershocks (considered as
point processes) according to the following distribution in time and space
\begin{equation}
\phi_m (r,t)~ dr~ dt = K~ 10^{\alpha_a m}
~\frac{\theta~c^{\theta}~dt}{(t+c)^{1+\theta}} ~
\frac{\mu~d^{\mu}~dr}{(r+d)^{1+\mu}}~,
\label{nmgjedl}
\end{equation}
where $r$ is the spatial distance to the main event. 
The spatial regularization distance $d$ accounts for
the finite rupture size. The power law kernel in space with
exponent $\mu$ quantifies the fact that
the distribution of distances between pairs of events is
well described by a power-law \cite{KJ1}. The ETAS model assumes that
each primary aftershock may trigger its own
aftershocks (secondary events) according to the same law, the secondary
aftershocks themselves may trigger tertiary aftershocks and so on,
creating a cascade process. The exponent $1+\theta$ is not the observable
Omori exponent $p_a$ but defines the ``bare'' Omori law for the
aftershocks of first generation.
The whole series of aftershocks, integrated over the whole space,
can be shown to lead to a ``renormalized'' (or dressed)
Omori law, which is the total observable Omori law \cite{JGR1Helmsor}.
To prevent the process from dying out, a small
Poisson rate of uncorrelated seismicity driven by plate tectonics 
is added to represent the effect of the tectonic
loading in earthquake nucleation.

The ETAS model predicts the following properties.
\begin{enumerate}
\item The same productivity law (\ref{formrho}) is found for the total 
number of events (and not just from the first generation) \cite{JGR1Helmsor}. 
This law describes an average power law increase of the number of directly
and indirectly triggered earthquakes per triggering earthquake as a function
of the mainshock size $s$.
\item The ETAS model finds a global direct Omori law (for 
aftershocks) different from the ``bare'' Omori
law $\sim 1/(t+c)^{1+\theta}$ defined in (\ref{nmgjedl}) for the first
generation aftershocks, with a renormalized exponent $p_a$ smaller than
from $1+\theta$ \cite{JGR1Helmsor}.
\item An inverse Omori law for foreshocks is found to result simply from
the existence of the direct ``bare'' Omori law (\ref{nmgjedl}) and from 
cascades of multiple triggering \cite{Helmfore}.
\item The ETAS model predicts
a modification of the magnitude distribution (\ref{gojfwo}) before a mainshock
of magnitude $m_M$,
characterized by an increase of the proportion of large earthquakes according
to the following expression \cite{Helmfore}:
\be
P(m|m_M) = (1-Q(t)) P_0(m) + Q(t) dP(m)~,
\label{GRaalaw}
\ee
where $P_0(m)$ is the unconditional GR distribution (\ref{gojfwo}),
\be
dP(m) \sim 10^{-b' (m-m_d)}~, ~~~~{\rm with}~~b'=b-\alpha_a~,
\label{gnnlala}
\ee
and
\be
Q(t) =  {C \over (t_c-t)^{\nu}}~,~~~~{\rm 
with}~~\nu={\theta(b-\alpha_a) \over \alpha_a}~,
\label{gnnlala2}
\ee
$t_c$ is the time of the mainshock and $C$ is a numerical constant.
The prediction (\ref{GRaalaw}) with (\ref{gnnlala},\ref{gnnlala2})
is that the magnitude distribution is modified upon the approach
of a mainshock by developing a bump in its tail which takes
the form of a growing additive power law contribution with a new $b'$-value.
This prediction has been validated very clearly by numerical simulations
of the ETAS model  \cite{Helmfore}.

\item In the ETAS model, the properties of foreshocks are independent
of the mainshock magnitude, because the magnitude of each event
is not predictable but is given by the GR law with a constant $b$-value \cite{Helmfore}.

\item By the mechanism of cascades of triggering, the ETAS model
also predicts the possibility for large distance and long-time build-up of foreshock
activity as well as the migration of foreshocks toward mainshocks.
\cite{HOS03,HSforedata03}. 

\end{enumerate}
These predictions of the ETAS
model are in good agreement with observations of the seismicity
in Southern California \cite{HSforedata03}.

\subsection{The critical earthquake model (CEM) \label{sakka}}

Maybe the first work on accelerated seismicity leading to the concept of criticality
is \cite{KBMalinovskaya}, who observed that the trailing total sum of the source 
areas of medium size earthquakes accelerates with time on the approach to a large or
great earthquake. 
The theoretical ancestor of the critical earthquake concept can probably be
traced back to \cite{Verejones}, who used a branching model to illustrate a cascade of
earthquake ruptures culminating in complete collapse interpreted as a great one.
Ref.~\cite{Allegre82} proposed a renormalization group analysis of a percolation model
of damage/rupture prior to an earthquake paralleling \cite{Reynolds},
which emphasized the critical point nature of earthquake rupture following an
inverse cascade from small to large scales. 
Refs.~\cite{Voight1,Voight2} are probably the first ones to introduce
the idea of a time-to-failure analysis in the form of a second order
nonlinear differential equation, which for certain values of the
parameters leads to a solution of the form of a time-to-failure
equation describing the power law acceleration of an observable with time:
\be
\epsilon(t) \propto A - B (t_c-t)^z~,
\label{mgmel}
\ee
where $\epsilon(t)$ is for instance the cumulative Benioff strain 
(square root of earthquake energy), $A$ and $B$ are positive constants,
$t_c$ is the critical time of the mainshock and $0<z<1$ is a critical exponent.
Ref.~\cite{Bufe} introduced equation (\ref{mgmel}) to fit and predict
large earthquakes. Their justification of (\ref{mgmel}) was a mechanical
model of material damage. Ref.~\cite{Bufe} did not mention the critical
earthquake concept. Ref.~\cite{SS95} proposed to reinterpret \cite{Bufe} and all these 
previous works and to generalize them using a statistical physics framework. 
The concept of a critical earthquake described in Ref.~\cite{SS95} corresponds to 
viewing  a major or great earthquake as a genuine critical point in the
statistical physics sense. In a nutshell, a critical point is characterized
by long-range correlations and by power laws describing the
behavior of various observables on the approach to the critical point.
This concept has been elaborated in subsequent works
\cite{SS90,Huangetla98,Bowman,JS99,Ouilrock,Moraplace,Zoller1,Zoller2,Zoller3,SS02}.
The critical earthquake model predicts a power-law increase of the number, 
of the cumulative displacement and of the average energy of foreshocks before 
large earthquakes. According to this model, the modification of seismic 
activity should be more apparent before larger mainshocks. Therefore, we should 
measure a positive value for $\alpha_f$ characterizing foreshocks.
The critical model also predicts that the preparation zone or
foreshocks cluster size $R_f$ should increase with the mainshock size
$s$ as  $R_f \sim s^{q_f}$ with  $q_f>0$, as observed by \cite{Bowman,KBMalinovskaya} 
(see \cite{JS99} for an extended compilation and discussion).

\section{Properties of the synthetic seismicity generated by the OFC model 
\label{result}}

\subsection{Definitions of foreshocks and aftershocks \label{fgjnkws}}

The first striking observation, extending the discovery of \cite{StephanPRL},
is that all the properties of real seismicity discussed in section \ref{nghjej}
are found to exist at least at a qualitative level in synthetic catalogs 
generated by the OFC model, as we now show.

Deciding what is a foreshock or an aftershock is not straightforward,
and several methods have been proposed to define aftershocks or foreshocks
\cite{Molchan92,HOS03}. 
A clear identification of foreshocks, aftershocks and mainshocks is
hindered by the fact that nothing distinguishes them in their seismic
signatures: at the present level of resolution of seismic inversions,
they are found to have the same double-couple structure and the same 
radiation patterns \cite{Houghjones}.
It is thus important to consider several alternative definitions.

\noindent
$\bullet$ {\it Definition} $d=0$. 
The common definition of foreshocks and aftershocks is as follows.
A mainshock is first defined as the largest event in a sequence. Foreshocks
and aftershocks are then defined as earthquakes in a pre-specified space-time 
domain around the mainshock, and are thus constrained to be
smaller than the mainshock. The space-time window used for foreshock and 
aftershock selection is defined such that the seismicity rate is much larger
than the background seismicity. 
For simplicity, we choose a space window that covers the whole lattice,
and we choose a time window with a duration equal to 1\% of the recurrence 
time of the mainshock, so that the seismicity within this window
is much larger than the background seismicity.

However, recent empirical and theoretical studies suggest that
this definition might be arbitrary and physically artificial
\cite{KK817,Shaw,Jones95,Houghjones,Felzer,HelmPRL,HSforedata03}.
Indeed, the magnitude of an earthquake seems to be unpredictable \cite{HelmPRL},
therefore the same mechanisms responsible for the triggering of small
earthquakes (usual ``aftershocks'' for $d=0$) may also explain the triggering of
larger earthquakes (defined as ``mainshocks'' for $d=0$). This hypothesis 
has been tested  in \cite{HSforedata03} and accounts very well for the 
properties of earthquake clustering in real seismicity. 
We thus use two other definitions of foreshocks and aftershocks, which 
do not constrain aftershocks and foreshocks to be smaller than the mainshock,
in order to test how the selection procedure impacts on 
foreshock and aftershock properties.

\noindent
$\bullet$ {\it Definition} $d=1$. 
Mainshocks are now defined as any earthquake that was not preceded by 
a larger earthquake within a time window equal to 1\% of the 
mainshock recurrence time, but can be followed by a larger event.
This rule aims to select as aftershocks the events that have been triggered 
directly or indirectly by the mainshock, removing the influence of large
earthquakes that occurred before the mainshock.
We use the same space-time domain as for $d=0$ to select foreshocks
and aftershocks around the mainshock.
Foreshocks are thus smaller than mainshocks but aftershocks can be larger.

\noindent
$\bullet$ {\it Definition} $d=2$. Same as $d=1$ without any constraint
on the aftershocks and mainshocks: each event of appropriate size
is a mainshock, preceded by foreshocks and aftershocks which are selected
solely on the basis of their belonging to the pre-specified 
space-time domain around the mainshock, with arbitrary magnitudes. 
For foreshocks, this corresponds to the ``Type II'' foreshocks introduced in
 \cite{Helmfore,HSforedata03} in order to remove any spurious dependence of 
foreshock properties on the mainshock magnitude.

For $d=2$, each earthquake can be a foreshock or an 
aftershock to several mainshocks and can also be a mainshock.

\subsection{Direct (aftershocks) and inverse (foreshocks) Omori laws \label{papf}}

\begin{figure}
\begin{center}
\includegraphics[width=8cm]{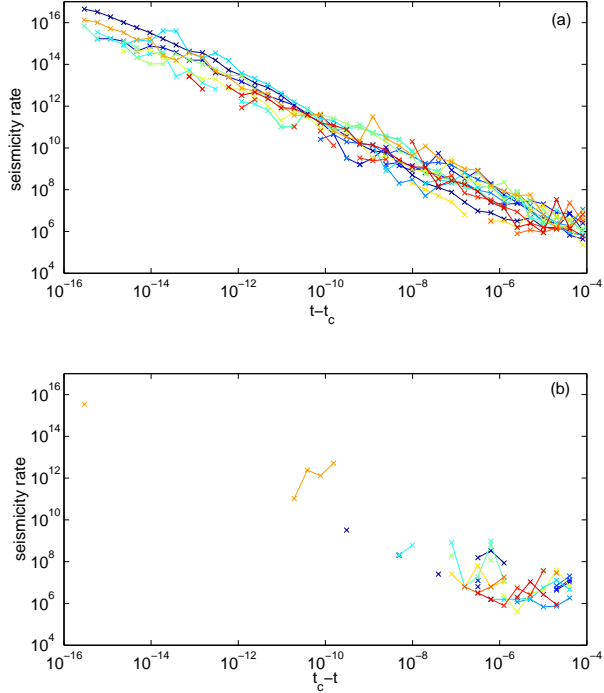}
\caption{\label{NAFtd0S2048p01}  $15$ individual sequences of 
aftershocks (a) and foreshocks (b), for mainshocks of size $s>2048$
with more than $1000$ foreshocks or aftershocks,
generated in a system of size $L=2048$, dissipation index $k=2$ and
selected with definition $d=0$.
}
\end{center}
\end{figure}

Figure \ref{NAFtd0S2048p01} shows $15$ individual sequences of foreshocks (bottom)
and aftershocks (top), for mainshocks of size $s>2048$
with more than $1000$ foreshocks or aftershocks,
generated in a system of size $L=2048$, dissipation index $k=2$ and
selected with definition $d=0$. The direct Omori law (\ref{first})
is strikingly clear for each individual sequence while,
for foreshocks, there is almost no increase
of the seismic rate for individual sequences: the inverse Omori law 
(\ref{firstsad}) is only observed when stacking many sequences, like in the ETAS model 
\cite{Helmfore}.
This is the clearest indication that the foreshock activity may be better
described by a cascade ETAS-type model than by a critical earthquake model
(but see below for other observations that may modulate this 
preliminary conclusion).
The OFC model shares this property with
empirical seismicity \cite{HSforedata03} and with the ETAS model 
\cite{Helmfore}.

\begin{figure}
\begin{center}
\includegraphics[width=8cm]{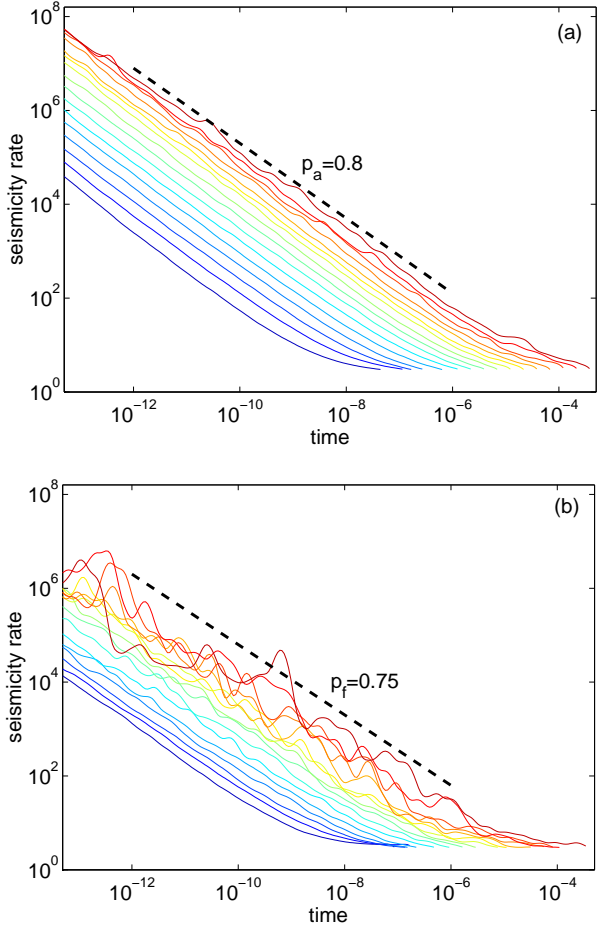}
\caption{\label{figominv} Direct ((a), aftershocks) and inverse
((b), foreshocks) Omori law for synthetic
catalogs generated with the OFC model with $L=2048$, dissipation 
index $k=2$ and definition $d=0$. The seismicity rate is normalized by
the background rate and by the number of mainshocks in each class.
The mainshock size increases from $s=2$ (bottom curve)
 to $s=2^{16}$ (top curve) with a factor 2 between each  curve.
}
\end{center}
\end{figure}

We now describe the results obtained by averaging over a large number of sequences,
which allow us to decrease the noise level and to look at smaller mainshocks. 
We have generated synthetic catalogs with the OFC model, using a lattice size
$L=128,256,512,1024,2048$ and different values of the dissipation index $k=0.5,1,2,4$.
For each catalog, we have selected aftershocks and foreshocks following the 
different definitions $d=0,1,2$ explained in section \ref{fgjnkws}.
We have then stacked all sequences by superposing all sequences
translated in time so that the mainshock occurs at time $t=0$.
We have first analyzed the change of the seismicity rate before and after
a mainshock.
For each range of mainshock size $s$ between 2 and $2^{16}$, increasing by a 
factor 2 between each class, we compute the average seismicity rate
$N_f(t)$ and $N_a(t)$ as a function of the time before and after the mainshock.
The results for $L=2048$, $k=2$ and $d=0$ are illustrated in Figure \ref{figominv}.
The rate of aftershocks obeys the direct Omori law (\ref{first}) and the 
 increase of the seismicity rate observed when averaging over many sequences
follows the inverse Omori law (\ref{firstsad}).
We measure the Omori exponents $p_f$ and $p_a$ by fitting the number
$N_a(t)$ of aftershocks and the number $N_f(t)$ of foreshocks by a power 
law $\sim 1/|t|^{p_{a,f}}$ using a linear regression of $\ln (N)$ as a function 
of $\ln |t|$, in the time interval $|t|>5 \cdot 10^{-14}$ and
$|t|<t_{max}$, where the upper bound $t_{max}$ is given by the condition that
the seismicity rate is much larger than the background level.

The duration of aftershock or foreshock sequences is much smaller than
the average recurrence time of earthquakes. This is also shown in
Figure 1 of Ref.~\cite{StephanPRL}, which represents the distribution
of inter-event times. A deviation from the Poisson distribution is
evident only for small times compared to the average inter-event times.

Table \ref{tab1} provides the values of the exponents $p_a$ and $p_f$
as a function of $L$, $k$ and $d$.
We find similar Omori exponents for foreshocks and aftershocks with
$p_f  \leq  p_a <1$. We observe the same time-dependence of the seismicity rate 
(same exponents  $p_a$ and $p_f$) for all mainshock sizes, only the absolute 
value of the seismicity rate depends on the mainshock magnitude.
The exponents $p_a$ and $p_f$ are found to increase with
 $k$ from $p_a \approx 0.5$ for $k=0.5$ to $p_a \approx 0.9$ for $k=4$, but
the duration of the aftershock and foreshock sequence does not change 
significantly with $k$. The Omori exponents do not depend on the rules
of selection $d$.
The number of foreshocks and aftershocks thus increases if the dissipation 
increases, and is almost negligible in the non-dissipative case.

\subsection{Dependence of the number of aftershocks and foreshocks 
with the mainshock size \label{alpha}}

\begin{figure}
\begin{center}
\includegraphics[width=8cm]{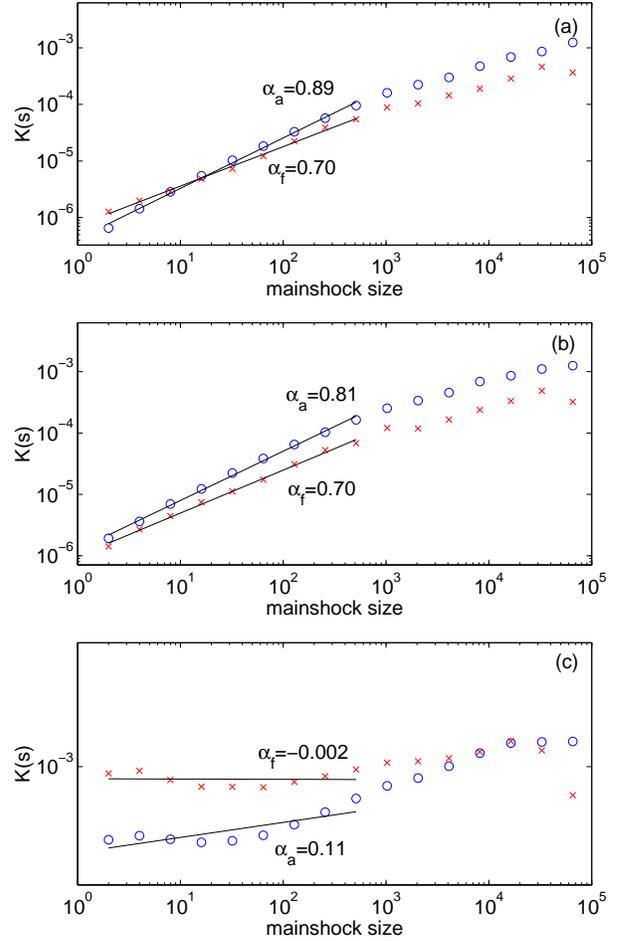}
\caption{\label{figalpha}
Productivity of aftershocks $K_a$ (circles) and foreshocks $K_f$
(crosses) as a function of the mainshock size $s$ for synthetic
catalogs generated with the OFC model with $L=2048$, dissipation 
index $k=2$ and definition $d=0$ (a), $d=1$ (b) and $d=2$ (c). 
}
\end{center}
\end{figure}

Figure \ref{figalpha} represents the dependence of the number of
aftershocks and foreshocks with the mainshock size, for different
rules of selection $d=0,1,2$.
We observe a power-law increase of the number of foreshocks $K_f$ and aftershocks
$K_a$  defined in (\ref{first}) and (\ref{firstsad}) with the mainshock size 
$s$ according to
\be
K_a(s) \sim s^{\alpha_a} ~,~~~~{\rm and}~~~~  K_f(s) \sim s^{\alpha_f}~,
\label{jgkkew}
\ee
for $s<10^3$.
The exponents  $\alpha_a$ and $\alpha_f$ measured for $s<1000$ 
increase with the dissipation index $k$
(see Table 1). The results are very similar for $d=1$ and $d=0$.
The exponent $\alpha_a$ is slightly smaller for $d=1$ than for $d=0$, because
for $d=0$ we impose the aftershocks to be smaller than mainshocks. 
Small events are more likely to trigger an event larger than
themselves than larger mainshocks and thus to be rejected from the
analysis. Therefore the rule $d=0$ underestimates the number 
of earthquakes triggered by small mainshocks.

For $d=2$ and for small $s$, $K_a(s)$ and $K_f(s)$ are much larger 
than with $d=0$, and are almost independent of $s$ for $s<100$.
This results from the fact that, for $d=2$ a significant fraction of
``mainshocks'' are triggered by a previous larger event, and thus
the events classified as aftershocks may be in fact triggered by
earthquakes that occurred before their ``mainshock''.
The results obtained with $d=2$ recovers those obtained with $d=0$ 
and $d=1$ for large $s$. The correct value of the exponent $\alpha_a$
of the aftershocks productivity law is thus the value obtained for $d=1$.

The  number of foreshocks is generally smaller than the number of aftershocks
and increases more slowly with $s$ ($\alpha_f \leq \alpha_a$).

For large mainshock sizes $s>1000$, we observe a saturation of $K_a$ and $K_f$,
and the numbers of foreshocks and aftershocks increase slower with $s$ than
predicted by (\ref{jgkkew}).
This saturation size does not depend neither on $k$ nor on $L$.
The effect of the system size $L$ is only to change the shape of the 
functional form of $K_a(s)$ and $K_f(s)$ for large $s$: the saturation of
$K_{a,f}(s)$ for $s>1000$ is more obvious for smaller $L$.

\subsection{Spatial distribution of foreshocks and aftershocks \label{q}}

\begin{figure}
\begin{center}
\includegraphics[width=12cm,angle=-90]{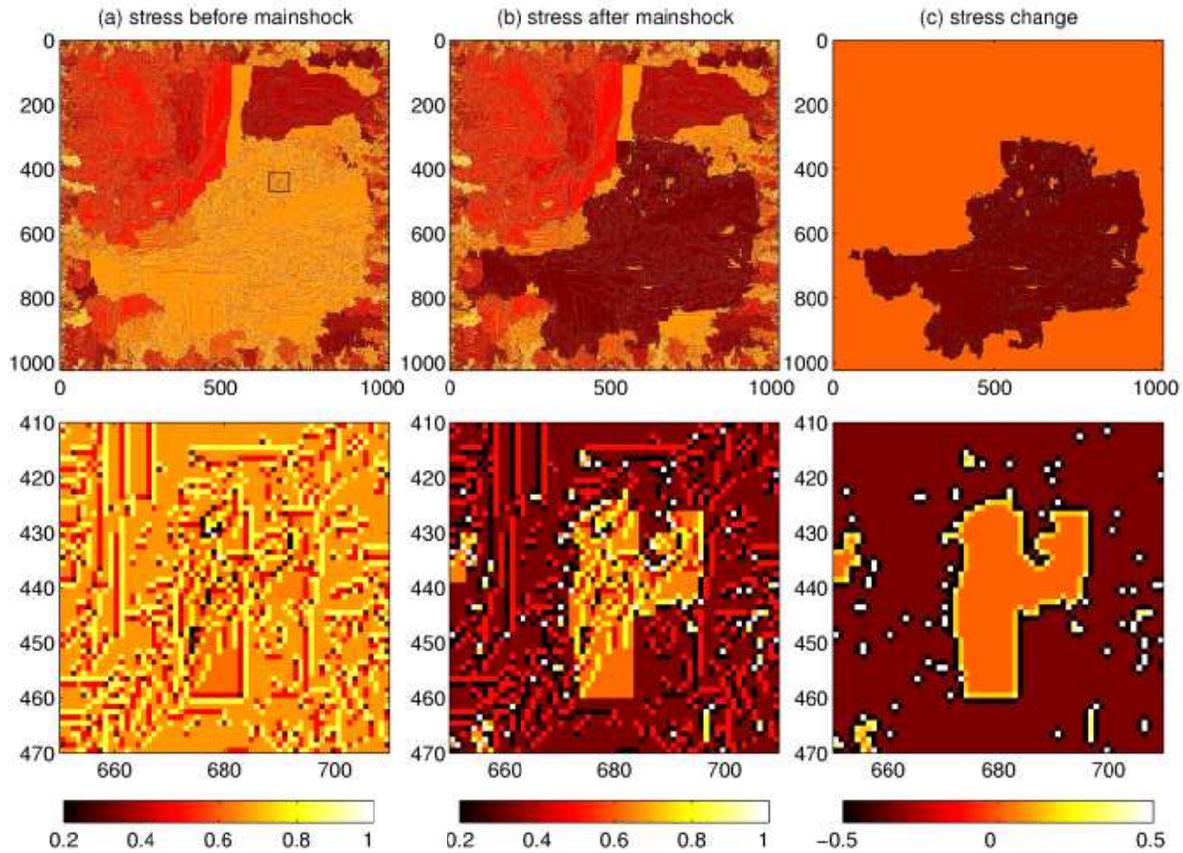}
\caption{\label{stressmap}
Stress field immediately before (a) and after (b) a mainshock.
The stress change due to the mainshock in shown in (c).
The elements that broke during the avalanche are shown in
dark in (c) (stress decrease), and were mostly close to the rupture
threshold before the mainshock (light gray in (a)).
The upper panels show the whole grid of size $L=1024$ and the lower
plots represent a subset of the grid delineated by the square in the
upper plot.}
\end{center}
\end{figure}

Figure \ref{stressmap} shows the stress field
immediately before and after a large mainshock. Following the
mainshock, many elements on the boundary of the avalanche and within
the avalanche have been loaded by the rupture, and are likely to 
generate aftershocks after the mainshock. There are a few large
patches of elements within the avalanche that did not break during the
mainshock, as illustrated in the lower panels of Figure \ref{stressmap},
but most aftershocks initiate on smaller clusters of a few unbroken elements
shown as white spots in the
central lower panel of Figure \ref{stressmap}. The density of such
white spots observed in this square is typical of the rest of the 
stress field over the area spanned by the mainshock. In the language
of seismicity, such white spots are ``asperities'' which carry a
large stress after the mainshock and are nucleation point for future aftershocks. 
These white spots are also found on the avalanche boundaries.

\begin{figure}
\begin{center}
\includegraphics[width=8cm]{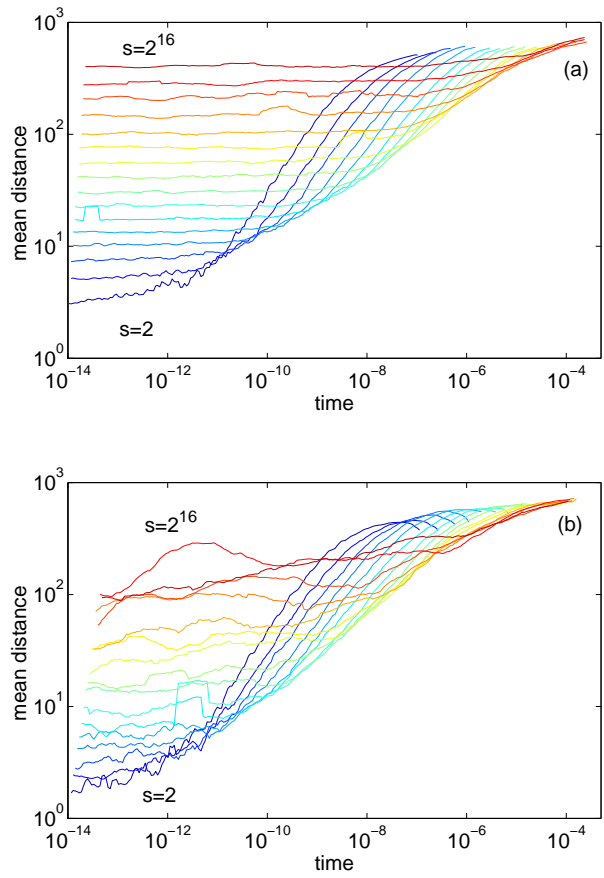}
\caption{\label{figr1}
Average distance between mainshocks and aftershocks (a) and
mainshocks and foreshocks (b) as a function of the time
after (a) or before (b) the mainshock, for synthetic
catalogs generated with the OFC model with $L=2048$, dissipation 
index $k=2$ and definition $d=0$. 
Each curve corresponds to a different mainshock size 
 size increasing from $s=2$ (bottom curve)
 to $s=2^{16}$ (top curve) with a factor 2 between each curve.}
\end{center}
\end{figure}

The aftershock cluster size $R_a$, defined as the
average distance between the mainshock and its aftershocks, is close 
to the mainshock size $\ell=\sqrt{s}$ at small times, and then increases 
to a value close to the system size $L$ at large times (see Figure \ref{figr1}). 
This crossover of $R_a(t)$ corresponds 
to a transition between the aftershock activity at small times and short
distances from (or within the area spanned by) the mainshock, 
to the uncorrelated seismicity at larger times
after the mainshock and with a uniform spatial distribution.
We observe the same pattern for the spatial distribution of foreshocks.
The time of the crossover of $R_a(t)$ and $R_f(t)$ increases with $s$
because the number of foreshocks and aftershocks increases with $s$ and
thus remain for a longer time above the uncorrelated background seismicity.
In the short times regime where uncorrelated seismicity is negligible, we find 
a very weak, if any, diffusion of aftershocks, as measured by 
an increase $R_a \sim t^{H_a}$ of the aftershock zone size with the time
after the mainshock. Similarly, we observe a very weak migration of 
foreshocks toward the mainshock, characterized by a decrease 
$R_f \sim (t_c-t)^{H_f}$ of the foreshock zone as the mainshock approaches.
The Hurst exponents $H_a$ and $H_f$ are not statistically different from $0$, 
showing that the size of the foreshock and aftershock zones do not change 
significantly with time. 

\begin{figure}
\begin{center}
\includegraphics[width=8cm]{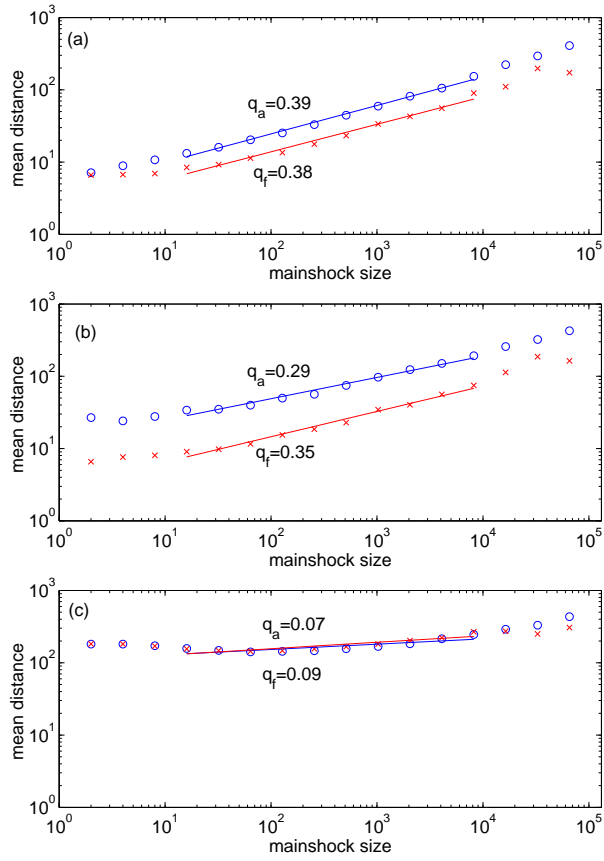}
\caption{\label{figr2} 
Average distances $R_a$ (circles) and $R_f$ (crosses) as a function of
the mainshock size $s$, for $k=2$, $L=2048$, and for definitions
$d=0$ (a), $d=1$ (b) and $d=2$ (c).
The  continuous lines are fits of $R_{a,f}(s)$ by a power-law $s^q$.}
\end{center}
\end{figure}

We observe on Figure \ref{figr2} a power-law increase of aftershock
zone size and of the foreshock zone size with the mainshock size according to
\be
R_a(s) \sim s^{q_a} ~,~~~~{\rm and}~~~~  R_f(s) \sim s^{q_f}~,
\label{gjkkwsl}
\ee
in the range $10<s<10^4$.
The exponents $q_a$ and $q_f$ are given in Table 1 as a function of $k$, $L$, 
and $d$. In contrast with real seismicity, the aftershock zone area is not 
proportional to the mainshock size $s^{0.5}$, but it increases slower with 
$s$ ($q_a \approx 0.3<0.5$ for $d=0$ or 1).
This is probably due to the effect of secondary aftershocks, which increase 
the effective size of the aftershock zone for small mainshocks. 
Secondary aftershocks are more important for $d=1$ than for $d=0$, 
which explains why $q_a$ is smaller for $d=1$ than for $d=0$.
The average value of the foreshock zone (``zone of mainshock preparation'')
is smaller than the aftershock zone, except for definition $d=2$.

An increase of $R_f$ with $s$ according to (\ref{gjkkwsl}) has been reported by
Ref.~\cite{KBMalinovskaya,Bowman} for individual sequences, with an exponent 
$q_f =0.44$ \cite{Bowman}, but was not observed by Ref.~\cite{HSforedata03} when using
stacked sequences and when allowing foreshocks to be larger that the mainshock ($d=2$).
This increase of $R_f$ with $s$ is not observed by the ETAS model, because
the magnitude of each earthquake is drawn at random, independently of previous 
seismicity, and thus all properties of foreshocks must be independent of the 
mainshock size. 

For $d=2$, the average zone sizes $R_a$ and of $R_f$ are much larger than for $d=0$,
because we include foreshocks and aftershocks larger than the mainshock.
>For $s<1000$, the values of $R_a$ and $R_f$ are almost constant of the order
of $R_{a,f} \approx 100$.
This may reflect the fact that, for $d=2$ and for small mainshocks, most 
aftershocks are not triggered by the mainshock but by a previous larger event.

\subsection{Distribution of avalanche sizes \label{b}}

For the whole catalogs, the distribution of event sizes is 
a power-law characterized by an exponent $b$, with an exponential roll-off
for large sizes $s>10^{-3} L^2$.
Table \ref{tab1} shows that the exponent $b$ increases as the dissipation
index increases, from $b \approx 0.7$ for $k=0.5$ to the realistic value $b 
\approx 0.95$ for $k=4$, in agreement with previous works
\cite{Lise,Grassberger}. 

\begin{figure}
\begin{center}
\includegraphics[width=16cm]{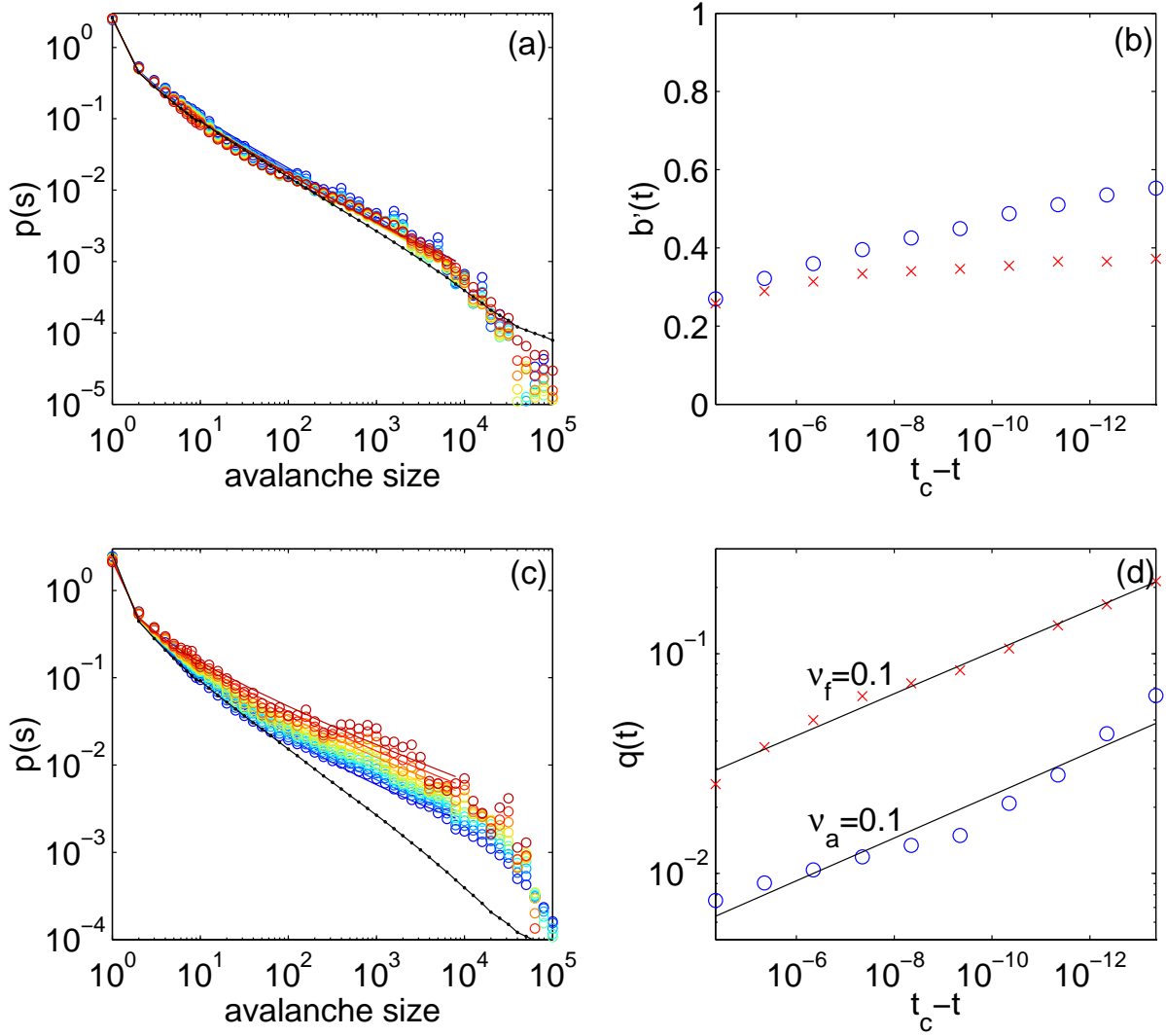}
\caption{\label{figpmaf1} 
Aftershock (a) and foreshock  (c) size distribution as a function of the time
before or after the mainshock, for $k=2$ and $L=2048$, and for a
mainshock size in the range $s=2048-4096$.
Aftershocks were selected with $d=1$ and foreshocks with $d=2$
(without constrains on the magnitude of foreshocks and aftershocks).
The avalanche size distributions are constructed with a logarithmic binning 
(linear bin in magnitudes), whose slope gives the GR $b$-value.
The black line in (a) in (c) shows the size distribution
for the whole catalog for reference.
The different colors from blue to red correspond to different time 
windows closer and closer to the mainshock, from $|t_c-t|=5 \times
10^{-5}$ to  $|t_c-t|=5 \times 10^{-14}$.
The colored lines in (a) and (c) are the fit of the foreshock and aftershock
magnitude distribution $P(m)$ (with $m=\log_{10}(s)$ as defined in 
(\ref{mgmlee})) by expression (\ref{GRaalaw}).
The corresponding values of $b'(t)$  (\ref{gnnlala}) and $Q(t)$
(\ref{gnnlala2}) are shown in (b) and
(d) for foreshocks (crosses) and aftershocks (circles),
as a function of time before/after the mainshock.}
\end{center}
\end{figure}

Figures \ref{figpmaf1} test the stationarity of the magnitude 
distribution for foreshocks and aftershocks.
The magnitude distributions for foreshocks and aftershocks have been 
fitted with expression (\ref{GRaalaw}) and these fits are shown as the 
colored lines.
The deviation of the magnitude distribution from the average GR law
for foreshocks and aftershocks are well
represented by (\ref{GRaalaw}) with $b'$ in the range
$0.3-0.5$ and with $Q(t)$ increasing as a power law according to 
(\ref{gnnlala2}) with $\nu \approx 0.1$ for foreshocks and for
aftershocks.

For aftershocks, the time-dependence of $Q(t)$ describes a decrease 
of the deviation from the GR law for latter aftershocks. 
While these fits are good, there is an important caveat: the
prediction $b'=b-\alpha$ of the ETAS model is not valid here,  as the OFC
model gives $\alpha_a \approx 0.8$ for $k=2$, which is slightly larger than
$b=0.78$. This regime $\alpha_a > b$ gives an explosive behavior in the
ETAS model with a stochastic finite time singularity \cite{Sorhelm02}
unless the GR law is truncated by an upper bound or is rolling over to a 
faster decay for the large earthquake sizes. 

\begin{figure}
\begin{center}
\includegraphics[width=8cm]{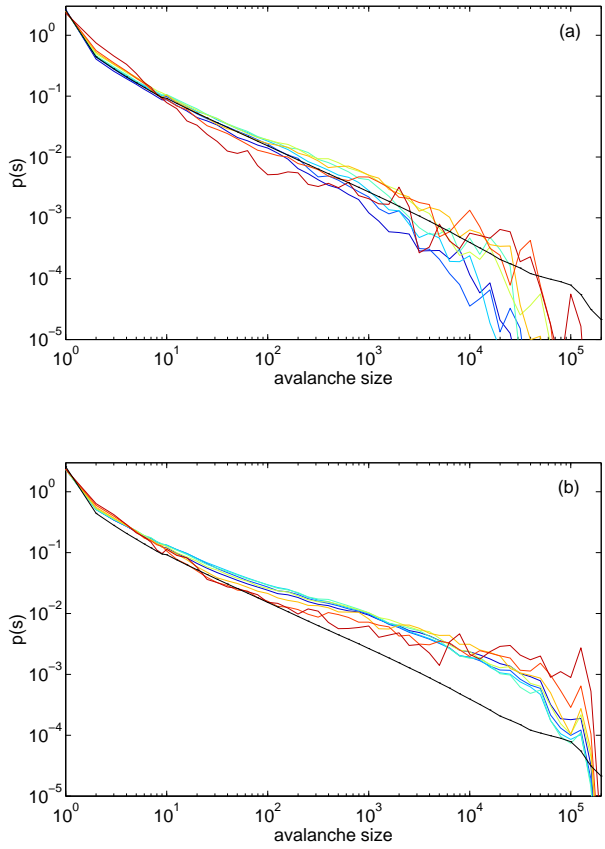}
\caption{\label{figpmaf2} 
Aftershock (a)  and foreshock (b) size distribution  for $k=2$ and $L=2048$.
Aftershocks ($d=1$) and foreshocks ($d=2$) are selected in a time
window $5\times 10^{-11}<|t_c-t|<5\times 10^{-10}$. Each curve
corresponds to a different mainshock size between $s=4$ (blue curve) 
and $s=2^{16}$ (red curve) increasing by a factor 4 between each
curve. The black line shows the (unconditional) distribution of the whole catalog.
}
\end{center}
\end{figure}

The modification of the magnitude distribution for aftershocks in the OFC model, 
is much weaker than for foreshocks, but is significant. 
This implies that the magnitude distribution in the OFC model is not stationary 
because the magnitude of triggered earthquakes is correlated with the 
mainshock magnitude, in contradiction with a crucial hypothesis
of the ETAS model and with real catalogs. 
Figure \ref{figpmaf2} shows that,
for $d=1$ (no constraint on aftershock magnitudes), the 
change of the magnitude distribution is almost independent on the mainshock 
magnitude (like in the ETAS model). However, there is a larger proportion of 
medium-size events for smaller mainshocks than for larger ones. This means that 
smaller mainshocks have a tendency to trigger smaller
aftershocks than larger mainshocks. 

This result is in contradiction with the ETAS model, which does not reproduce
a dependence of the aftershock magnitude as a function of the size of the triggering 
earthquake or as a function of the time since the mainshock.
Observations of real seismicity \cite{HelmPRL} do not show any dependence
between the aftershock magnitude and the mainshock magnitude, but the catalogs 
available are much smaller than for our OFC simulations.

\section{Mechanisms for foreshocks and aftershocks in the OFC model
\label{discu}}

Is the increase of the number of aftershocks and foreshocks with the
mainshock magnitude real or is it just the result
of a selection bias introduced by the standard definition $d=0$, which requires
that mainshocks are the largest events in the cluster?
Indeed, in the OFC catalogs, the clear power-law increase of the number 
of aftershocks and foreshocks with the size $s$ of the mainshock, found when 
we define the mainshock as the largest event ($d=0$), almost
disappears when aftershocks or foreshocks are not constrained to be
smaller than the mainshock ($d=2$), as shown in Figure \ref{figalpha}. 
The question of the impact
of the definition is thus essential.

For foreshocks, we consider two possible interpretations, the ETAS model 
described in section \ref{gjkdks} and  the critical earthquake model (CEM) 
summarized in section \ref{sakka}.
The fact that, using $d=0$, both the number of foreshocks and the
average foreshock cluster size increase with the mainshock size $s$
seems to favor the critical model, but Ref.~\cite{Helmfore} have shown
that the constraint
that foreshocks must be smaller than the mainshock ($d=0$) leads to an
artificial increase of the number of foreshocks and of the
foreshock cluster size with $s$ (see Fig. 5 of \cite{Helmfore}).
However, this spurious increase of the number of foreshocks with the mainshock
magnitude should be observed only for small $s$, and should
not exist in the case $d=2$ (without constraints on foreshock and mainshock 
magnitudes).
The fact that a weak increase of the number of foreshocks with $s$
is observed even for very large $s$ and for all definitions $d=0,1,2$
suggests that the effect is genuine.
Such a dependence of foreshock properties on the mainshock size
cannot be reproduced with the ETAS model, but suggest that the critical
model provides a more relevant description of these observations.

The case $d=2$ destroys the dependence of the aftershock number 
with $s$ for small $s$, which is a real physical property of aftershocks
because it is also observed for $d=1$. 
The same effect may also be at work for foreshocks and explain why,
if the properties of foreshocks are physically dependent on
the mainshock magnitude, this dependence is not observed
when using $d=2$.  Definition $d=2$ may be mixing ``critical foreshocks'',
i.e. events that belong to the preparation phase of the future mainshock,
with ``triggering-triggered'' pairs (which are the usual ``foreshock-mainshock''
pairs in the ETAS model).  Thus, the absence of conditioning  ($d=2$) 
seems to destroy the dependence of both
the foreshock and aftershock properties with the mainshock magnitudes, i.e.,
$\alpha_{a,f} \approx 0$ and $q_{a,f} \approx 0$ both for foreshocks 
and aftershocks. But, for aftershocks, we know for a truth that the dependence 
of the number ($\alpha_a$) and the cluster size ($q_a$) are real and physical 
because they are observed in the case $d=1$ which do not constrain aftershocks 
to be smaller than the mainshocks.

For aftershocks selected according to $d=2$ and for small mainshock
sizes $s$, the scenario, according to which
most aftershocks are not triggered by the mainshock per se but by a
previous larger event, seems
to explain both the fact that (i) the number of aftershocks
is almost constant with $s$ for $s<100$ and (ii) the size of the
aftershock cluster is almost constant with $s$ for small $s$.
It is however surprising, if this interpretation is correct, that
we observe a pure Omori law for aftershocks and foreshocks in the 
case $d=2$, without
any change in the Omori exponent with $s$ and without any roll-off at small times.
 
The observations that earthquake clustering in the OFC model is much weaker than
for real seismicity implies that the ``branching ratio''
(average number of triggered earthquakes of first generation per
mother-earthquake) must be close to $0$ for the ETAS model to match
the OFC model. But such a small branching ratio would lead to Omori's
exponent larger than $1$, in contradiction with our simulations. 
Thus, the pure branching model of triggered seismicity is insufficient
to fully account for the quantitative aspect of the OFC seismicity.

In order to better understand the mechanism responsible for aftershocks 
in the OFC model, we have imposed several types of perturbations to the 
normal course of an OFC simulation
to obtain the response of the system. First, we have simulated 
random isolated disturbances consisting in choosing randomly and
independently $1024$ sites and
adding to them random amounts of stress drawn in the interval 
$[0, 0.01]$ or $[0, 0.1]$. Repeating such disturbance 100000 times,
we find no observable seismicity triggered by this perturbation.
This shows that the aftershocks require a coherent spatial organization 
over a broad area. In contrast, pasting (or grafting) the stress map 
as shown in figure \ref{stressmap} of those sites 
that participated in a given large
mainshock in a given simulation and their nearest neighbors onto 
another independent simulation gives a perfect Omori law following this graft
in this second simulation as if the foreign stress map of the first
simulation was part of the 
second simulation. The resulting
aftershock sequences are similar to natural sequences.		
This demonstrates that the presence of asperities close to the failure
threshold inside the boundary of the avalanche can produce 
realistic aftershock sequences with an Omori law temporal decay.
To investigate further if it
is the spatial connectivity of the perturbation which is important to get
aftershocks, we have also raised simultaneously the stress by various amounts within
squares of size $64\times 64$. Performing this perturbation 40000 times at random instants,
we observe that this sometimes results in quite large earthquakes immediately
following the perturbation, but there is no significant activity afterwards, 
and nothing that looks like an Omori law of aftershocks. Thus, it seems 
impossible to generate aftershocks by introducing a
random or deterministic spatially extended perturbation. This suggests
that aftershocks require not only the 
occurrence of mainshocks to redistribute stress
but also a spatial organization of the stress field prior to the mainshock
so that many ``asperities'' (as illustrated in figure \ref{stressmap}) can
be created by the interplay of the stress organization before the mainshock
and the stress redistribution by the mainshock.

\section{Conclusion \label{conclu}}

The results obtained in this paper can be viewed from two different
perspectives. On the one hand, we are adding to the phenomenology of
one of the most studied model of self-organized criticality. 
Extending Hergarten and Neugebauer's announcement \cite{StephanPRL}, we
have shown evidence that the self-organized critical state of the OFC
model is much richer than previously thought, with important
correlation patterns in space and time between avalanches. We have 
obtained quantitative scaling laws describing 
the spatio-temporal clustering of events in the OFC model.
On the other hand, we have shown that what is probably the simplest
possible mechanism for the generation of earthquakes (slow tectonic loading and sudden
stress relaxation with local stress redistribution) is sufficient to 
recover essentially all known properties documented in seismic catalogs, at
a qualitative level. Specifically, we have found that foreshocks and aftershocks
follow the inverse and direct Omori power laws, as in real seismicity
but with smaller exponents. 
We have found the productivity power law of aftershocks
as in real seismicity. In contrast with real seismicity, we also found
a power-law increase of foreshock productivity with the mainshock size.
The nucleation of
aftershocks at ``asperities'' located on the mainshock rupture plane 
or on the boundary of the avalanche is also
in agreement with seismic observations. These findings are interesting because
they add on the list of possible mechanisms for foreshocks and aftershocks that
have been discussed previously in the literature.

We also found that the predictability increases with the mainshock magnitude
in the OFC model, because the number of foreshocks seems to increase
with the magnitude of the mainshock, a feature that 
is not observed in real seismicity \cite{HSforedata03}.
See the debate in Nature at http://helix.nature.com/debates/earthquake/ and 
\cite{PC94,Geller} for related discussions on the predictability of 
real earthquakes and the use of models from statistical physics.

While most of the OFC dynamics can be qualitatively captured by the simple
ETAS model of triggered seismicity, this property that the number
of foreshocks seems to increase with the magnitude of the mainshock
is better explained by the critical
earthquake model. We have systematically compared the statistical properties
of the avalanches generated by the dynamics of the OFC model with those
predicted by the ETAS model on the one hand and by the critical earthquake model
on the other hand. These two models constitute end-member models of seismicity.
We can conclude that, while the
detailed quantitative predictions of the ETAS model for
its application to the OFC model are not exactly right, the
physical mechanism of cascades of triggered events inherent in the ETAS
model captures at least qualitatively the observed regularities found in the OFC model.
This suggests a picture in which future avalanches are triggered by past avalanches
through ``asperities'' located either within the plane of past avalanches or at their
boundaries. In addition, we find that this triggering mechanism presents
a degree of cooperativity,
as the number of foreshocks increases with the mainshock size.
In other words, this suggests that asperities interact via avalanches, and 
when their number and size increase in a given location, they can produce larger avalanches.

\acknowledgments
This work was partially supported by NSF-EAR02-30429, by
the Southern California Earthquake Center (SCEC) and
by the James S. Mc Donnell Foundation 21st
century scientist award/studying complex system.
SCEC is funded by NSF Cooperative Agreement EAR-0106924 and USGS Cooperative
Agreement 02HQAG0008.  The SCEC contribution number for this paper is ***.

{}

\clearpage
\begin{table}
\caption{
Aftershock and foreshock properties in the OFC model as a function of
the system size $L$ and of the dissipation index $k$, for different
definitions of foreshocks and aftershocks $d$ (see section \ref{fgjnkws}).
$b$ is the exponent of the cumulative distribution of avalanche sizes,
measured for the whole catalog and discussed in section \ref{b}.
 $p_a$ and $p_f$ are the exponents of the direct and inverse Omori laws 
(see section \ref{papf}).
$\alpha_a$ and $\alpha_f$ characterize the dependence of the aftershock
and foreshock rate with the mainshock size $s$ (see section \ref{alpha}). 
$q_a$ and $q_f$ describe the scaling of the aftershocks and foreshock 
zone size with $s$ (see section \ref{q}).
$N_{\rm max}$ is the maximum rate of aftershocks at $t=10^{-14}$
normalized by the background rate. }
\begin{center}
\begin{tabular}{lcccccccccc}
\hline
$k$   & $L$    & $b$    & $d$ & $p_a$ & $p_f$ & $\alpha_a$ & 
$\alpha_f$ & $q_a$ & $q_f$ & $N_{\rm max}$  \\
\hline
$0.5$ & $512$  & $0.71$ & $0$ &	$0.5$ & $0.5$ &	$-0.2$      & $-0.3$ 
&	     &	      & $10^3$  \\
$0.5$ & $512$  &        & $1$ &	$0.5$ & $0.5$ &	$-0.2$      & 
&	     &	      & $10^3$  \\
$0.5$ & $512$  &        & $2$ &	$0.5$ & $0.5$ &	$-0.3$      & $-0.5$ 
&	     &	      & $10^3$  \\
\hline
$0.5$ & $1024$ & $0.67$ & $0$ &	$0.6$ & $0.6$ &	$0.06$      & $-0.29$ 
&	     &	      & $10^3$  \\
$0.5$ & $1024$ &        & $1$ &	$0.6$ & $0.6$ &	$0.05$      & $-0.21$ 
&	     &	      & $10^3$  \\
$0.5$ & $1024$ &        & $2$ &	$0.5$ & $0.5$ &	$-0.42$     & $-0.21$ 
&	     &	      & $10^3$  \\
\hline
$1$   & $512$  & $0.76$ & $0$ &	$0.65$& $0.65$&	$0.63$      & $0.36$ 
& $$      & $$     & $10^6$  \\
$1$   & $512$  &        & $1$ &	$0.65$& $0.65$&	$0.6$       & $0.31$ 
& $$      & $$     & $10^6$  \\
$1$   & $512$  &        & $2$ &	$0.65$& $0.65$&	$0.11$      & $-0.1$ 
& $$      & $$     & $10^6$  \\
\hline
$1$   & $1024$ & $0.73$ & $0$ &	$0.65$& $0.65$&	$0.63$      & $0.35$ 
& $0.20$  & $0.17$ & $10^6$  \\
$1$   & $1024$ &        & $1$ &	$0.65$& $0.65$&	$0.58$      & $0.34$ 
& $0.19$  &$ 0.13$ & $10^6$  \\
$1$   & $1024$ &        & $2$ &	$0.65$& $0.65$&	$0.11$      & $0.09$ 
& $0.11$  &$ 0.14$ & $10^6$  \\
\hline
$2$   & $128$ & $0.80$ & $0$ &	$0.65$& $0.65$&	$0.52$      & $0.27$ 
& $0.30$  & $0.44$ & $10^6$  \\
$2$   & $128$ &        & $1$ &	$0.65$& $0.65$&	$0.41$      & $0.21$ 
& $0.29$  & $0.18$ & $10^5$  \\
$2$   & $128$ &        & $2$ &	$0.60$& $0.60$&	$0.16$      & $-0.1$ 
& $0.27$  & $0.01$ & $10^5$  \\
\hline
$2$   & $256$ & $0.81$ & $0$ &	$0.75$& $0.75$&	$0.77$      & $0.57$ 
& $0.35$  & $0.29$ & $10^7$  \\
$2$   & $256$ &        & $1$ &	$0.70$& $0.70$&	$0.67$      & $0.52$ 
& $0.31$  & $0.26$ & $10^7$  \\
$2$   & $256$ &        & $2$ &	$0.70$& $0.70$&	$0.21$      & $0.01$ 
& $0.21$  & $0.03$ & $10^7$  \\
\hline
$2$   & $512$ & $0.80$ & $0$ &	$0.75$& $0.75$&	$0.82$      & $0.64$ 
& $0.37$  & $0.37$ & $10^8$  \\
$2$   & $512$ &        & $1$ &	$0.75$& $0.75$&	$0.75$      & $0.61$ 
& $0.31$  & $0.32$ & $10^8$  \\
$2$   & $512$ &        & $2$ &	$0.75$& $0.75$&	$0.16$      & $0.01$ 
& $0.13$  & $0.06$ & $10^8$  \\
\hline
$2$   & $1024$& $0.78$ & $0$ &	$0.80$& $0.80$&	$0.88$      & $0.68$ 
& $0.36$  & $0.30$ & $10^8$  \\
$2$   & $1024$&        & $1$ &	$0.80$& $0.80$&	$0.80$      & $0.67$ 
& $0.28$  & $0.30$ & $10^8$  \\
$2$   & $1024$&        & $2$ &	$0.75$& $0.75$&	$0.14$      & $0.03$ 
& $0.08$  & $0.07$ & $10^8$  \\
\hline
$2$   & $2048$& $0.76$ & $0$ &	$0.80$& $0.75$&	$0.89$      & $0.70$ 
& $0.39$  & $0.36$ & $10^8$  \\
$2$   & $2048$&        & $1$ &	$0.81$& $0.75$&	$0.80$      & $0.70$ 
& $0.29$  & $0.35$ & $10^8$  \\
$2$   & $2048$&        & $2$ &	$0.80$& $0.75$&	$0.11$      & $0.00$ 
& $0.07$  & $0.09$ & $10^8$  \\
\hline
$4$   & $512$& $0.95$ & $0$ &	$0.85$& $0.80$&	$0.90$      & $0.71$ 
& $0.34$  & $0.40$ & $10^9$  \\
$4$   & $512$&        & $1$ &	$0.85$& $0.80$&	$0.77$      & $0.66$ 
& $0.23$  & $0.39$ & $10^9$  \\
$4$   & $512$&        & $2$ &	$0.80$& $0.80$&	$0.22$      & $0.07$ 
& $0.10$  & $0.19$ & $10^9$  \\
\hline
$4$   & $1024$& $0.92$ & $0$ &	$0.85$& $0.80$&	$0.91$      & $0.72$ 
& $0.36$  & $0.33$ & $10^9$  \\
$4$   & $1024$&        & $1$ &	$0.80$& $0.80$&	$0.78$      & $0.66$ 
& $0.26$  & $0.30$ & $10^9$  \\
$4$   & $1024$&        & $2$ &	$0.80$& $0.80$&	$0.19$      & $0.06$ 
& $0.13$  & $0.07$ & $10^9$  \\
\\
\hline
\label{tab1}
\end{tabular}
\end{center}
\end{table}

\clearpage

\clearpage

\clearpage

\clearpage

\clearpage

\end{document}